\documentclass[aip,jmp,reprint,onecolumn]{revtex4-1}

\usepackage{amssymb}
\usepackage{amsfonts}
\usepackage{amstext}
\usepackage{graphicx}
\usepackage{amscd}
\usepackage{amsmath}

\newcommand{\ihbar}{\imath \hbar}

\newcommand{\Pe}{\mathbb{P}e}

\newcommand{\E}{\mathbb{E}}
\newcommand{\Ran}{\mathrm{Ran}}

\begin{document}

\title{Geometric phases in quantum control disturbed by classical stochastic processes}

\author{David Viennot}
\affiliation{Institut UTINAM (CNRS UMR 6213, Universit\'e de Franche-Comt\'e), 41bis Avenue de l'Observatoire, BP 1615, 25010 Besan\c con cedex, France.}

\begin{abstract}
We describe the geometric (Berry) phases arising when some quantum systems are driven by control classical parameters but also by outer classical stochastic processes (as for example classical noises). The total geometric phase is then divided into an usual geometric phase associated with the control parameters and a second geometric phase associated with the stochastic processes. The geometric structure in which these geometric phases take place is a composite bundle (and not an usual principal bundle), which is explicitely built in this paper. We explain why the composite bundle structure is the more natural framework to study this problem. Finally we treat a very simple example of a two level atom driven by a phase modulated laser field with a phase instability described by a gaussian white noise. In particular we compute the average geometric phase issued from the noise.
\end{abstract}

\maketitle

\section{Introduction}
The quantum control can be modelled by quantum systems driven by parameter dependent selfadjoint Hamiltonians $H(x)$, where $x$ is a set of classical parameters -- the control parameters -- spanning a manifold $M$. A quantum control problem consists then to find a path $\mathcal C$ in $M$ such that the quantum evolution generated by $t \mapsto H(x(t))$ (where $t \mapsto x(t)$ is the time parametrization of $\mathcal C$) induces an evolution of the wave function from an initial determined state to a wanted target state. Since the control parameters are classical, they generally evolve slowly with respect to the quantum proper time of the system. An adiabatic theorem can then be applied \cite{messiah,nenciu}. Let $\{\lambda_a(x)\}_a$ be the eigenvalues of $H(x)$, the strict adiabatic theorem \cite{messiah} states that: if $\psi(0) = |a,x(0)\rangle$ ($\lambda_a$ is supposed non-degenerate and $|a,x\rangle$ is its associated eigenvector) then $\psi(t) = e^{-\ihbar^{-1} \int_0^t \lambda_a(x(t'))dt'} e^{- \int_{\mathcal C} A} |a,x(t) \rangle$ ($\forall t>0$), with $A = \langle a,x|d_M|a,x \rangle$ where $d_M$ is the exterior differential of $M$. $e^{- \int_{\mathcal C} A}$ is called the geometric phase (or the Berry phase) \cite{berry}. For a degenerate eigenvalue we have $\psi(t) = e^{-\ihbar^{-1} \int_0^t \lambda_a(x(t'))dt'} \sum_{j=1}^{n_a} \left[ \Pe^{- \int_{\mathcal C} A} \right]_{ji} |a(j),x(t) \rangle$ with $\psi(0) = |a(i),x(0)\rangle$ ($\{|a(j),x\rangle\}_{j=1,...,n_a}$ is a set of continuous orthonormal eigenvectors associated with $\lambda_a$) and
\begin{equation}
A(x) = \left( \begin{array}{ccc} \langle a(1),x|d_M|a(1),x\rangle & ... & \langle a(1),x|d_M|a(n_a),x\rangle \\ \vdots & \ddots & \vdots \\ \langle a(n_a),x|d_M|a(1),x\rangle & ... & \langle a(n_a),x|d_M|a(n_a),x\rangle \end{array} \right) \in \Omega^1(M,\mathfrak g)
\end{equation}
$\Pe$ denotes the path ordered exponential \cite{nakahara}, $\Omega^1(M,\mathfrak g)$ denotes the set of $\mathfrak g$-valued differential 1-form of $M$ ($\mathfrak g = \mathfrak u(n_A)$ is the set of order $n_a$ antiselfadjoint matrices). $\Pe^{- \int_{\mathcal C} A}$ is the non-abelian geometric phase \cite{wilczek}. The geometric phases are mathematically described by a principal bundle \cite{simon} over $M$, $P(M,G) \xrightarrow{\pi_P} M$, with structure group $G = U(n_a)$ (the set of order $n_a$ unitary matrices). The geometric phase represents the result of the control in addition to the simple deformation of the instantaneous eigenvectors. In particular, for a closed path $\mathcal C$ (a closed control) the geometric phase $\Pe^{- \oint_{\mathcal C} A}$ is the single effect at the end of the control (it is then a holonomy in the principal bundle).\\
In realistic situations, the experimental control apparatus is not perfect, and in practice the control parameters can be affected by noise. More generally, some other classical stochastic processes can arise during the control. The goal of this paper is the description of the geometric phases in this situation. Some works about geometric phases subject to noise have been already published, which concern the effect of the noise on the global dynamics \cite{gaitan,li} or the perturbative effect of a small noise \cite{chiara}. The present paper adopt a different way. We want consider situations where the stochastic processes can be included in the geometric framework and are then not considered only as global dynamical effects. We want consider the geometric phase arising from these stochastic processes whithout a perturbative assumption. To this, we use a natural geometric framework to define these geometric phases, the composite bundle theory \cite{sardanashvily,viennot1,viennot2}.\\
This paper is organized as follows. Next section presents the model of quantum systems driven by control parameters and stochastic processes considered in this paper. It is devoted to the construction of the composite bundle used in the theory. The naturality of this structure to model the present problem is discussed. Last section presents a simple example in order to illustrate the concepts introduced in this paper, i.e. a two level atom diven by a phase modulated laser field subject to noise.

\section{The model and the composite bundle}
We consider a quantum system described by an Hilbert space $\mathcal H$ and driven by a set of classical control parameters $x$ which belong to a manifold $M$. The quantum system is described by a selfadjoint Hamiltonian $H:M \to \mathcal H$ ($H$ is a continuous map from the control manifold $M$ to the Hilbert space $\mathcal H$ which is independent from $M$). The quantum system is also subject to a set of classical stochastic parameters $y$ which belong to a manifold $R$. For a control $t \mapsto x(t)$, the random functions $y(t)$ are solutions of a stochastic differential equation (SDE):
\begin{equation}
\label{SDE}
\dot y^\mu = f^\mu(x(t),y) + k^{\mu \nu}(x(t),y) \eta_\nu(t)
\end{equation}
where $\{\eta_\mu\}$ are stochastic random variables.\\
Let $\Gamma(M,S)$ be the set of the solutions of the SDE eq.(\ref{SDE}). We can see $\Gamma(M,S)$ as the set of the sections of a fibre bundle $S(M,R) \xrightarrow{\pi_S} M$ with base manifold $M$ and typical fibre $R$. We denote by $S$ its total manifold and by $\pi_S$ its projection. Let $\{U^\alpha \}_\alpha$ be an atlas of $M$ (an open cover of $M$). We denote by $\chi^\alpha : U^\alpha \times R \to S$ the local trivialization of $S$ over $U^\alpha$.\\

We suppose that the selfadjoint Hamiltonian governing the dynamics of the quantum system can be written as
\begin{equation}
H_+(s) = W(s) H(\pi_S(s)) W(s)^{-1}
\end{equation}
where $s\in S$ is a point of $S$. $s$ represents a configuration of the stochastic and control parameters accessible via a stochatic evolution described by the SDE eq.(\ref{SDE}). $s\in S$ can be represented by local coordinates $(x,y) = {\chi^{\alpha}}^{-1}(s)$ where $x$ are the independent variables and $y$ are the dependent variables. $H(x)$ is the Hamiltonian governing the control without disturbance and $W(s) \in \mathcal U(\mathcal H)$ is an unitary operator of the Hibert space which represents the disturbance induced by the stochastic processes. We note that this form includes cases where the stochastic processes induce a perturbation $\epsilon V(s)$ with $\epsilon \ll 1$. Indeed let $W(s) = e^{\imath \epsilon w(s)}$ where $w(s)$ is a selfadjoint operator. We have then $H_+(s) = H(\pi_S(s)) + \epsilon [\imath w(s),H(\pi_S(s))] + \mathcal O(\epsilon^2)$. Finally we set $V(s) = [\imath w(s),H(\pi_S(s))]$ (in practice it can be difficult to solve this equation to find $w(s)$).\\

Let $\lambda_a(x)$ be a $n_a$-fold degenerate eigenvalue of $H(x)$ which does not cross another eigenvalue and let $\{|a(j),x \rangle \}_{j=1,...,n_a}$ be an associated set of orthonormal eigenvectors. $\lambda_a$ is a continuous real function on $M$. $\lambda_a(\pi_S(s))$ is an eigenvalue of $H_+(s)$ with the eigenvectors $\{|a(j),s\rangle_+ = W(s)|a(j), \pi_S(s) \rangle\}_{j=1,...,n_a}$.\\

Let $G = U(n_a)$ be the group of order $n_a$ unitary matrices. We denote by $P(M,G) \xrightarrow{\pi_P} M$ the $G$-principal bundle of the geometric phases associated with the quantum control without disturbance. This bundle is defined as follows. It is generally impossible to define a set of continuous eigenvectors on the whole of $M$ with the same convention. We have then only locally defined continuous eigenvectors $U^\alpha \ni x \mapsto |a(j),x \rangle^\alpha$, with
\begin{equation}
\forall x \in U^\alpha \cap U^\beta, \qquad |a(j),x \rangle^\beta = \sum_{k=1}^n [g^{\alpha \beta}_P(x)]_{jk} |a(k),x \rangle^\alpha
\end{equation}
The change basis matrices $g^{\alpha \beta}(x) \in G$ constitute the transition functions of $P$ and define then $P$. We denote by $\phi^\alpha_P : U^\alpha \times G \to P$ its local trivialization over $U^\alpha$.\\
The definition of $P$ can be also expressed by using the eigenequation. Let $P_a(x) \in \mathcal B(\mathcal H)$ be the orthogonal eigenprojector associated with $\lambda_a(x)$ ($\mathcal B(\mathcal H)$ denotes the set of bounded operators of $\mathcal H$): $[H(x),P_a(x)]=0$ with $P_a(x)^2 = P_a(x)$ and $P_a(x)^\dagger = P_a(x)$. We define a vector bundle $E_a = \bigsqcup_{x \in M} \Ran P_a(x)$ over $M$ ($\sqcup$ denotes the disjoint union). Since $\lambda_a(x)$ does not cross another eigenvalue, $P_a(x)$ has a constant rank (equal to $n_a$) on the whole of $M$. $E_a$ is then a well defined locally trivial vector bundle on $M$ and the index $a$ is globally defined. We can note that if all eigenvalues of $H$ are isolated we have
\begin{equation}
\bigoplus_{b} E_b = M \times \mathcal H
\end{equation}
where we have supposed that $H(x)$ has a pure point spectrum. Nevertheless, this assumption is too strong and we need only for the present discussion that the considered eigenvalue $\lambda_a(x)$ is isolated on the whole of $M$. The adiabatic assumption states that the wave function remains projected onto $\Ran P_a(x(t))$ for a slow evolution $t\mapsto x(t)$. The typical fiber of $E_a$ is $\mathbb C^{n_a}$ and its local trivializations are defined by
\begin{equation}
\phi^\alpha_{E_a} : \begin{array}{rcl} U^\alpha \times \mathbb C^{n_a} & \to & {E_a}_{|\pi^{-1}_{E_a}(U^\alpha)} \\ \left(x,\left(\begin{array}{c} c_1 \\ \vdots \\ c_{n_a} \end{array} \right) \right) & \mapsto & \sum_{i=1}^{n_a} c_i |a(i),x \rangle^\alpha \end{array}
\end{equation} 
The eigenstates $|a(i),x\rangle^\alpha \in \mathcal H$ can be viewed as local sections of $E_a$. They are orthonormal with respect to the scalar product of $\mathcal H$ ($E_a$ is not endowed with an inner product, in particular we have ${^\alpha}\langle a(i),x|a(j),x \rangle^\alpha = \delta_{ij}$ $\forall x \in M$ but for $x \not= x'$ we have ${^\alpha}\langle a(i),x|a(j),x' \rangle^\alpha \not= \delta_{ij} \delta(x-x')$).  Let $\rho: G \to \mathcal L(\mathbb C^{n_a})$ be the matrix representation of the Lie group $U(n_a)$ (which can be viewed as a representation of $G$ on $\mathbb C^n$). Let $\hat \rho : G \to E_a$ be the action of $G$ on $E_a$ induced by $\rho$:
\begin{eqnarray}
\forall g\in G, \quad \hat \rho(g) \sum_{i=1}^{n_a} c_i |a(i),x\rangle^\alpha & = & \sum_{i,j=1}^{n_a} \rho(g)_{ij} c_j |a(i),x \rangle^{\alpha} \nonumber \\
& = & \sum_{j=1}^{n_a} c_j \widetilde{|a(j),x\rangle^\alpha}
\end{eqnarray}
where $\{\widetilde{|a(j),x\rangle^\alpha}\}_{j=1,...,m}$ is another set of orthonormal eigenvectors of $\Ran P_a(x)$. The action of $G$ represents then a change of the basis of $\Ran P_a(x)$ in $\mathcal H$ and a change of the considered $n_a$ local sections of $E_a$. $E_a(M,\mathbb C^{n_a})$ is the vector bundle associated with $P(M,G)$ by using $\rho$, i.e. $E_a(M,\mathbb C^{n_a}) = P(M,G) \times_{\rho} \mathbb C^{n_a}$.\\
The different bundles can also be defined by using the theory of the standard and universal bundles (see \cite{rohlin}). Let $G_{n_a}(\mathcal H)$ be the manifold of rank $n_a$ orthogonal projectors of $\mathcal H$ ($G_{n_a}(\mathcal H)$ is called a complex Grassmannian manifold), and let $V_{n_a}(\mathcal H)$ be the manifold of orthonormal $n_a$-frame of $\mathcal H$ ($V_{n_a}(\mathcal H)$ is called a Stiefel manifold). $V_{n_a}(\mathcal H)$ can be viewed as a $G$-principal bundle over $G_{n_a}(\mathcal H)$ which is universal for the $U(n_a)$-principal bundles (in the sense of the fibre bundle universal classification theorem \cite{rohlin}). $P_a:M \to G_{n_a}(\mathcal H)$ is then a map from the control manifold $M$ to the Grassmannian manifold which induces $P(M,G)$, i.e. we have the following commutative diagram:
$$ \begin{CD}
P @>{\Pr_2}>> V_{n_a}(\mathcal H)\\
@V{\pi_P}VV @VV{\pi_V}V \\
M @>>{P_a}> G_{n_a}(\mathcal H)
\end{CD} $$
where $P = P_a^*V_{n_a}(\mathcal H)= \{(x,B) \in M \times V_{n_a}(\mathcal H) \text{ such that } P_a(x)=\pi_V(B)\}$ with $\Pr_2(x,B)=B$. Let $S_{n_a}(\mathcal H) = V_{n_a}(\mathcal H) \times_\rho \mathbb C^{n_a}$ be the vector bundle over $G_{n_a}(\mathcal H)$ associated with the universal bundle $V_{n_a}(\mathcal H) \to G_{n_a}(\mathcal H)$. $S_{n_a}(\mathcal H)$ is a Steenrod standard bundle \cite{rohlin}. $E_a$ is induced by $P_a$ from $S_{n_a}(\mathcal H)$ in a same manner than $P$.\\
The usual geometric phases are associated with the horizontal lifts in $P(M,G) \xrightarrow{\pi_P} M$ endowed with a connection $\omega_P \in \Omega^1(P,\mathfrak g)$ defined by the gauge potential $A^\alpha_P(x) \in \Omega^1(M,\mathfrak g)$ ($[A^\alpha_P(x)]_{jk} = {^\alpha}\langle a(j),x|d_M|a(k),x\rangle^\alpha$). We note that this connection is not flat. Let $Z^\alpha(x) \in \mathfrak{M}_{\dim \mathcal H \times n_a}(\mathbb C)$ be the matrix of the $n_a$-eigenvectors represented on an arbitrary orthonormal basis of $\mathcal H$ denoted $(|k\rangle)_{k=1,...,\dim \mathcal H}$ independent from $x$ (to simplify the discussion, we suppose that $\dim \mathcal H < + \infty$, but it can be easily generalized to the infinite dimensional case). 
\begin{equation}
Z^\alpha(x) = \left(\begin{array}{ccc} \langle 1|a(1),x\rangle^\alpha & ... &  \langle 1|a(n_a),x\rangle^\alpha \\ \vdots &  & \vdots \\  \langle \dim \mathcal H|a(1),x\rangle^\alpha & ... &  \langle \dim \mathcal H|a(n_a),x\rangle^\alpha \end{array} \right)
\end{equation}
where the $\langle.|.\rangle$ denotes the scalar product of $\mathcal H$. We have then
\begin{equation}
A_P^\alpha(x) = Z^\alpha(x)^\dagger d Z^\alpha(x) \in \Omega^1(M,\mathfrak g)
\end{equation}
where $\dagger$ denotes the matricial transconjugate. Except for the special cases where $n_a = \dim \mathcal H$ (and where $H(x) = \lambda_a(x) 1_{\mathcal H}$, $1_{\mathcal H}$ being the identity operator of $\mathcal H$), $Z^\alpha(x)^\dagger \not= Z^\alpha(x)^{-1}$ (moreover if $n_a < \dim \mathcal H$, $Z^\alpha(x)$ is not invertible since it is not a square matrix), and the curvature $F_P = dA_P^\alpha + A_P^\alpha \wedge A_P^\alpha = dA_P^\alpha + \frac{1}{2} [A_P^\alpha, A_P^\alpha]$ is not zero.\\
Naively we could think that the geometric phase for the disturbed control could be described in $P(M,G) \xrightarrow{\pi_P} M$ endowed with another connection $\tilde \omega_P = \omega_P + \delta \omega_P$ where $\delta \omega_P$ represents the deviations induced by the stochastic processes. The horizontality in $P$ would then be defined by $\ker \tilde \omega_P$ in place of $\ker \omega_P$. In fact this is not a natural viewpoint. The quantum system does not evolve parallely to $t \mapsto x(t)$ with an internal process which modifies its definition of the horizontality. In fact the stochastic processes modify the external environment felt by the quantum system. The quantum system reacts to its modified environment and not directly to the ideal control configuration $x$. The quantum evolution is then parallel to $t \mapsto s(t)$. The horizontality for the disturbed control must be then defined in a larger total space $P_+$ which includes also ``horizontal directions parallel to $R$''. The map $\pi_S : S \to M$ permits to naturally realize the enlargement of $P$, by defining $P_+$ as the bundle:
\begin{equation}
P_+ = \pi_S^* P = \{(s,p) \in S \times P \text{ such that } \pi_S(s) = \pi_P(p) \}
\end{equation}
$P_+ \xrightarrow{\Pr_2} P \xrightarrow{\pi_P} M$ (with $\Pr_2(s,p) = p$) constitutes a composite bundle representing the enlargement induced by the stochastic processes. But $P_+$ can be viewed not only as a bundle over $P$ but also as a $G$-bundle over $S$, $P_+(S,G) \xrightarrow{\pi_+} S$, with local trivializations defined by
\begin{equation}
\phi^\alpha_+ : \begin{array}{rcl} \pi_S^{-1}(U^\alpha) \times G & \to & {P_+}_{| \pi_S^{-1}(U^\alpha)} \\ (s,g) & \mapsto & \left(s, \phi^\alpha_P(\pi_P(s),g) \right) \end{array}
\end{equation}
$\pi_S$ being a continuous map, $\{\pi_S^{-1}(U^\alpha)\}_\alpha$ constitutes an atlas of $S$. By construction the transition functions of $P_+(S,G) \xrightarrow{\pi_+} S$ are $\forall s \in \pi_S^{-1}(U^\alpha \cap U^\beta)$, $g^{\alpha \beta}_+(s) = g^{\alpha \beta}_P(\pi_S(s))$. This is precisely the transition functions defined by $H_+$:
\begin{equation}
\forall s \in \pi_S^{-1}(U^\alpha \cap U^\beta), \qquad |a(j),s \rangle^\beta_+ = \sum_{k=1}^{n_a} [g^{\alpha \beta}_P(\pi_P(s))]_{jk} |a(k),s \rangle^\alpha_+
\end{equation}
We have then a second composite bundle $P_+ \xrightarrow{\pi_+} S \xrightarrow{\pi_S} M$. The first floor $S \to M$ represents the reaction of the environment to the control, and the second floor $P_+ \to S$ represents the reaction of the quantum system to the environment. This composite bundle represents then the fact that the ``control signal'' does not act directly on the system but ``goes through'' an intermediate ``medium'' $S$ which randomly distorts the ``control signal''. The comparison between disturbed and undisturbed controls can be realized by considering both the two ``parallel'' composite bundles, i.e. the following commutative diagram:
$$ \begin{CD}
P_+ @= P_+ \\
@V{\pi_+}VV @VV{\Pr_2}V \\
S & & P \\
@V{\pi_S}VV @VV{\pi_P}V \\
M @= M
\end{CD} $$
A description based on the two sets of parameters $(x,y)$ can be derived from $P_+ \xrightarrow{\pi_+} S \xrightarrow{\pi_S} M$. Let $\{P^\alpha_{++} \xrightarrow{\pi_{++}} U^\alpha \times R\}_\alpha$ be the set of local $G$-bundles defined by the following local trivializations
\begin{equation}
\phi^\alpha_{++}: \begin{array}{rcl} U^\alpha \times R \times G & \to & P^\alpha_{++} \subset P_+ \\ (x,y,g) & \mapsto & \phi^\alpha_+(\chi^\alpha(x,y),g) \end{array}
\end{equation}
The transition between two local bundles satisfies, for $x \in U^\alpha \cap U^\beta$, $y\in R$ and $g\in G$:
\begin{eqnarray}
\phi^\beta_{++}(x,y,g) & = & \phi^\beta_+(\chi^\beta(x,y),g) \\
& = & \phi^\alpha_+(\chi^\beta(x,y),g^{\alpha \beta}_P(x)g) \\
& = & \phi^\alpha_{++}(x,\varphi^{\alpha \beta}_x(y),g^{\alpha \beta}_P(x)g)
\end{eqnarray}
where the diffeomorphism $\varphi^{\alpha \beta}_x : R \to R$ is defined by $\varphi^{\alpha \beta}_x(y) = {\chi^\alpha}^{-1} \circ \chi^\beta(x,y)$. $\{P^\alpha_{++}\}_\alpha$ is a kind of twisted bundle \cite{viennot2,mackaay}. The torsion functions $\varphi^{\alpha \beta}$ are the obstruction to lift $\{P^\alpha_{++}\}_\alpha$ into a principal $G$-bundle over $M \times R$ ($\{P^\alpha_{++}\}_\alpha$ can be lift into a principal bundle if and only if the first floor of the composite bundle is trivial, i.e. if $S = M \times R$).\\

The $G$-bundle $P_+(S,G) \xrightarrow{\pi_+} S$ is endowed with a connection $\omega_+ \in \Omega^1(P_+,\mathfrak g)$. The connection $\omega_P \in \Omega^1(P,\mathfrak g)$ of $P(M,G) \xrightarrow{\pi_P} M$ can be pulled back into $P_+(S,G) \xrightarrow{\pi_+} S$ by the cotangent map $\Pr_2^*$: $\Pr_2^* \omega_P \in \Omega^1(P_+,\mathfrak g)$. The deviation induced by the stochastic processes can then be defined as $\delta \omega_+ = \omega_+ - \Pr_2^* \omega_P$.\\
Let $\sigma^\alpha \in \Gamma(\pi^{-1}_S(U^\alpha),P_+)$ be a local section of $P_+(S,G)$ and $\sigma^{\alpha *}$ be its pull-back. Let $\chi^{\alpha *}$ be the pull-back of $\chi^\alpha$. The gauge potential of the twisted bundle associated with the composite bundle $P_+ \to S \to M$ endowed with the connection $\omega_+$ is then $A_+ = \chi^{\alpha *} \sigma^{\alpha *} \omega_+ \in \Omega^1(U^\alpha \times R,\mathfrak g)$. It is defined by
\begin{equation}
A_+^\alpha(x,y) = \left( \begin{array}{ccc} {_+^\alpha}\langle a(1),\chi^\alpha(x,y)|d_{M\times R}|a(1),\chi^\alpha(x,y)\rangle_+^\alpha & ... & {_+^\alpha}\langle a(1),\chi^\alpha(x,y)|d_{M \times R}|a(n_a),\chi^\alpha(x,y)\rangle_+^\alpha \\ \vdots & \ddots & \vdots \\ {_+^\alpha}\langle a(n_a),\chi^\alpha(x,y)|d_{M \times R}|a(1),\chi^\alpha(x,y)\rangle_+^\alpha & ... & {_+^\alpha}\langle a(n_a),\chi^\alpha(x,y)|d_{M \times R}|a(n_a),\chi^\alpha(x,y)\rangle_+^\alpha \end{array} \right)
\end{equation}
We suppose that $\psi(0) = |a(i),\chi^\alpha(x(0),y(0)) \rangle^\alpha_+$ and we assume that the evolution is adiabatic for a control $t \mapsto x(t)$ (we do not discuss this assumption, the proof an adiabatic theorem with stochastic processes is not the subject of this paper). Let $\mathcal C_+$ be the path drawn by $(x(t),y(t))$ in $U^\alpha \times R$ (to simplify the discussion we suppose here that the path remains included in $U^\alpha$). The wave function is then at time $t>0$
\begin{equation}
\psi(t) = e^{- \ihbar^{-1} \int_0^t \lambda_a(x(t'))dt'} \sum_{j=1}^{n_a} \left[ \Pe^{- \int_{\mathcal C_+} A_+^\alpha} \right]_{ji} W(\chi^\alpha(x(t),y(t)))|a,x(t) \rangle^\alpha
\end{equation}
In accordance with the discussion concerning the composite bundles, the generator of the geometric phase can be divided in three terms:
\begin{itemize}
\item the generator of the geometric phase of the undisturbed quantum control: $A_P^\alpha = \chi^{\alpha*} \sigma^{\alpha *} \Pr_2^* \omega_P$
\begin{equation}
A_P^\alpha(x) = \left( \begin{array}{ccc} {^\alpha}\langle a(1),x|d_M|a(1),x\rangle^\alpha & ... & {^\alpha}\langle a(1),x|d_M|a(n_a),x\rangle^\alpha \\ \vdots & \ddots & \vdots \\ {^\alpha}\langle a(n_a),x|d_M|a(1),x\rangle^\alpha & ... & {^\alpha}\langle a(n_a),x|d_M|a(n_a),x\rangle^\alpha \end{array} \right)
\end{equation}
\item the generator of the geometric phase induced by the stochastic processes $\chi^{\alpha*} \sigma^{\alpha *} \delta \omega_+$ which can be splited into a generator measuring the effect of the stochastic parameters variations:
\begin{equation}
A_{Qx}^\alpha(y) = \left( \begin{array}{ccc} {^\alpha}\langle a(1),x|W^\alpha(x,y)^{-1}d_RW^\alpha(x,y)|a(1),x\rangle^\alpha & ... & {^\alpha}\langle a(1),x|W^\alpha(x,y)^{-1}d_RW^\alpha(x,y)|a(n_a),x\rangle^\alpha \\ \vdots & \ddots & \vdots \\ {^\alpha}\langle a(n_a),x|W^\alpha(x,y)^{-1}d_RW^\alpha(x,y)|a(1),x\rangle^\alpha & ... & {^\alpha}\langle a(n_a),x|W^\alpha(x,y)^{-1}d_RW^\alpha(x,y)|a(n_a),x\rangle^\alpha \end{array} \right)
\end{equation}
where $W^\alpha(x,y) = W(\chi^\alpha(x,y))$,
\item and a generator measuring the possible conjugate effects of the variations of the control and of the stochastic processes:
\begin{equation}
A_{Ty}^\alpha(x) = \left( \begin{array}{ccc} {^\alpha}\langle a(1),x|(W^\alpha(x,y)^{-1}d_MW^\alpha(x,y))|a(1),x\rangle^\alpha & ... & {^\alpha}\langle a(1),x|(W^\alpha(x,y)^{-1}d_MW^\alpha(x,y))|a(n_a),x\rangle^\alpha \\ \vdots & \ddots & \vdots \\ {^\alpha}\langle a(n_a),x|(W^\alpha(x,y)^{-1}d_MW^\alpha(x,y))|a(1),x\rangle^\alpha & ... & {^\alpha}\langle a(n_a),x|(W^\alpha(x,y)^{-1}d_MW^\alpha(x,y))|a(n_a),x\rangle^\alpha \end{array} \right)
\end{equation}
We can imagine that in practice this last term is small. In accordance with the adiabatic assumption, the variations of the control parameters are probably slow with respect to the stochastic fluctuations. We can then write this as ${^\alpha}\langle a,x|W^{\alpha-1} \frac{\partial W^\alpha}{\partial y^\nu}|a,x\rangle^\alpha \dot y^\nu \ll {^\alpha}\langle a,x|W^{\alpha-1} \frac{\partial W^\alpha}{\partial x^\mu} |a,x \rangle^\alpha \dot x^\mu $ permitting to neglect the contribution of $A_{Ty}$ with respect to the contribution of $A_{Qx}$.
\end{itemize}
The expression of the geometric phase when several charts of the atlas are crossed and the decomposition of the composite bundle holonomies into holonomies associated with $\omega_P$ and with $\delta \omega_+$ (and with an intertwining term) can be found in ref. \cite{viennot2} (the structure of the composite bundles considered in ref. \cite{viennot2} is a bit different from the structure considered in the present work -- the roles of $M$ and $R$ are reversed -- but the essential results are similar and can be easily adaptated).

\section{Example : two level atom driven by a phase modulated laser field with noise}
We consider a two-level atom driven by a laser field. The phase $\theta$ of the laser field is time modulated in order to realize the control. The laser phase is subject to stochastic fluctuations represented by $\delta \theta$. In the rotating wave approximation (RWA) with one photon, the Hamiltonian of the system is
\begin{equation}
H_+ = \frac{\hbar}{2} \left( \begin{array}{cc} 0 & \Omega e^{\imath (\theta+\delta \theta)} \\ \Omega e^{-\imath (\theta+\delta \theta)} & 2 \Delta \end{array} \right)
\end{equation}
where $\Omega = |\langle 1|\vec \mu \cdot \vec E |2 \rangle|$ ($\vec \mu$ being the atomic electric dipolar momentum, $\vec E$ is the intensity and the polarization of the laser field, and $\{|1\rangle,|2 \rangle\}$ are the states of the isolated atom). $\Delta$ is the detuning, i.e. $\hbar \Delta = E_2-E_1-\hbar \omega$ (where $E_1,E_2$ are the atomic energy levels and $\omega$ is the laser frequency). To simplify the discussion, these parameters are supposed constant. We suppose moreover that the phase fluctuations follow a Wiener process \cite{puri} described by the following SDE:
\begin{equation}
\label{wiener}
\delta \dot \theta = k(\theta(t)) \eta(t)
\end{equation}
where $\eta$ is a stochastic random variable and such that the function $k$ is $2\pi$-antiperiodic:
\begin{equation}
k(\theta + 2\pi) = - k(\theta)
\end{equation}
The control manifold $M$ describes the laser phase configurations and can be then assimilated to a circle $M = S^1$. We do not restrict the amplitudes of the phase fluctuations and we set $R = \mathbb R$. The solutions of the SDE eq. (\ref{wiener}) can be written as
\begin{equation}
\delta \theta(t) = \int_0^t k(\theta(t')) \eta(t')dt'
\end{equation}
The anti-periodicity of $k$ induces that
\begin{equation}
\delta \theta \xrightarrow{\theta \to \theta + 2 \pi} - \delta \theta
\end{equation}
Clearly, the manifold $S(S^1,\mathbb R)$ is an infinitely large Moebius strip \cite{nakahara}. Let $\{U^\alpha\}_{\alpha=1,2,3}$ be the atlas of $M=S^1$ defined by
\begin{eqnarray*}
\forall \theta \in U^1, & \quad & \ell^1(\theta) \in ]-\epsilon,\pi+\epsilon[ \\
\forall \theta \in U^2, & \quad & \ell^2(\theta) \in ]\pi-\epsilon,\frac{3\pi}{2}+\epsilon[ \\
\forall \theta \in U^3, & \quad & \ell^3(\theta) \in ]\frac{3\pi}{2}-\epsilon,2\pi+\epsilon[
\end{eqnarray*}
where $\epsilon \ll 1$ and $\{\ell^\alpha\}_\alpha$ are the coordinate functions associated with the atlas. Let $\chi^\alpha : U^\alpha \times \mathbb R \to S(S^1,\mathbb R)$ be the local trivializations of the Moebius strip and $\varphi^{\alpha \beta}_\theta = {\chi^\alpha_\theta}^{-1} \circ \chi^\beta_\theta$ be the torsion functions. We have
\begin{eqnarray*}
\forall \theta \in U^1 \cap U^2, & \quad & \ell^2(\theta)-\ell^1(\theta) = 0 \text{ and } \varphi^{12}_\theta(\delta \theta) = \delta \theta \\
\forall \theta \in U^2 \cap U^3, & \quad & \ell^3(\theta)-\ell^2(\theta) = 0 \text{ and } \varphi^{23}_\theta(\delta \theta) = \delta \theta \\
\forall \theta \in U^1 \cap U^3, & \quad & \ell^3(\theta)-\ell^1(\theta) = 2\pi \text{ and } \varphi^{13}_\theta(\delta \theta) = -\delta \theta
\end{eqnarray*}
The hamiltonian of the undisturbed control of the atom is
\begin{equation}
H(\theta) = \frac{\hbar}{2} \left( \begin{array}{cc} 0 & \Omega e^{\imath \theta} \\ \Omega e^{-\imath \theta} & 2 \Delta \end{array} \right)
\end{equation}
$r = \sqrt{\Omega^2+\Delta^2}$ is an eigenvalue of $H$ and its associated eigenvector is
\begin{equation}
|+,\theta \rangle = \frac{1}{\sqrt{2r(r+\Delta)}} \left(\begin{array}{c} \Omega e^{\imath \theta} \\ \Delta + r \end{array} \right)
\end{equation}
We have omited the chart indices since $|+,\theta \rangle$ is continuously defined on the whole of $M=S^1$, we have then $g^{\alpha \beta}_P(\theta) = 1$ and the bundle $P(M,G) \to M$ (with $G=U(1)$) is trivial: $P = S^1 \times U(1)$ (the manifold $P$ is a torus). The generator of the geometric phase of the undisturbed control is
\begin{equation}
A_P(\theta) = \langle +, \theta|d_M|+,\theta \rangle = \frac{\imath \Omega^2}{2r(r+\Delta)} d\theta
\end{equation}
The operator representing the disturbance effect is
\begin{equation}
W = \left( \begin{array}{cc} e^{- \frac{\imath}{2} \delta \theta} & 0 \\ 0 & e^{\frac{\imath}{2} \delta \theta} \end{array} \right)
\end{equation}
The generator of the geometric phase induced by the stochastic processes is then
\begin{equation}
A^\alpha_{Q\theta}(\delta \theta) = \langle +,\theta|W^{\alpha}(\delta \theta)^{-1} d_RW^\alpha(\delta \theta)|+,\theta \rangle = \imath \frac{\Delta}{r} d\delta \theta
\end{equation}
(In order to simplify the notation, we denote indifferently by $\delta \theta$ the Wiener process, the point of the manifold $R = \mathbb R$ and the coordinate of the point of $R = \mathbb R$).\\
Finally, for a control $t \mapsto \theta(t)$, if $\psi(0) = |+,\theta(0)\rangle$ (we suppose that $\delta \theta(0)=0$) we have
\begin{eqnarray}
\psi(t) & = & e^{- \ihbar^{-1} rt} e^{- \int_{\mathcal C} A_P} e^{- \int_{\mathcal S} A_Q} W(\delta(t))|+,\theta(t) \rangle \\
& = & e^{- \ihbar^{-1} rt} e^{- \imath \frac{\Omega^2}{2r(r+\Delta)} \theta(t)} e^{- \imath \frac{\Delta}{r} \delta \theta(t)} \left( \begin{array}{cc} e^{-\frac{\ihbar}{2} \delta \theta(t)} & 0 \\ 0 & e^{\frac{\imath}{2} \delta \theta(t)} \end{array} \right) |+,\theta(t) \rangle
\end{eqnarray}
where $t \mapsto \delta \theta(t)$ is a solution of the SDE eq.(\ref{wiener}), $\mathcal C$ is the path drawn by $\theta(t)$ in $S^1$ and $\mathcal S$ is the path drawn by $\delta \theta(t)$ in $\mathbb R$. The geometric phase induced by the stochastic processes is then $e^{- \imath \frac{\Delta}{r} \delta \theta(t)}$ (it arises in addition to the effect of $W$). By construction, after an instantaneous tranformation consisting to one turn around $S^1$, the wave function acquires a phase $e^{- \imath \frac{\Omega^2}{r(r+\Delta)} \pi + \imath \frac{2 \Delta}{r} \delta \theta_0}$ where $\delta \theta_0$ is the value of $\delta \theta$ at the begining of the turn.\\

We suppose now that $\eta$ is a Gaussian white noise \cite{puri}, i.e.
\begin{eqnarray}
\E[\eta(t)] & = & 0 \\
\E[\eta(t_1)\eta(t_2)] & = & D \delta(t_2-t_1)
\end{eqnarray}
where $D>0$ is a constant and $\E$ denotes the probability expectation. In that case we have (see ref. \cite{puri}) the following conditional probability
\begin{equation}
p(\delta \theta,t|\delta \theta_0,t_0) = \frac{1}{\sqrt{2 \pi K(t,t_0)}} e^{- \frac{(\delta \theta - \delta \theta_0)^2}{2 K(t,t_0)}}
\end{equation}
with
\begin{equation}
K(t,t_0) = D \int_{t_0}^t k(\theta(t'))^2 dt'
\end{equation}
We have then $\forall \alpha \in \mathbb R$
\begin{equation}
\E[e^{- \imath \alpha \delta \theta(t)}] =  e^{- \alpha^2 \frac{K(t,0)}{2}}
\end{equation}
We can then compute the expected wave function with respect to the stochastic process:
\begin{equation}
\E[\psi(t)] = e^{-\ihbar^{-1} rt} e^{-\imath \frac{\Omega^2}{2r(r+\Delta)} \theta(t)} \left( \begin{array}{cc} e^{-(\frac{\Delta}{r}+\frac{1}{2})^2 \frac{K(t,0)}{2}} & 0 \\ 0 & e^{-(\frac{\Delta}{r}-\frac{1}{2})^2 \frac{K(t,0)}{2}} \end{array} \right) |+,\theta(t) \rangle
\end{equation}
Let a bath of atoms in low temperature driven by a laser field such that each atom is submitted to an independent noise $\eta$. In that case $\E[\psi(t)]$ is the average wave function for the bath. It is interesting to remark that even if the adiabatic assumption is perfectly individually satisfied by each atom, $|\langle +,\theta(t)|W(\delta(t))^{-1}|\psi(t)\rangle| = 1$, the average population is $|\E[\langle +,\theta(t)|W(\delta(t))^{-1}|\psi(t)\rangle]|=e^{- \frac{\Delta^2}{r^2} \frac{K(t,0)}{2}}<1$. The study of the bath shows then an attenuation of the ``signal'' of the population ``+'' due to the geometric phase induced by the noises.

\section{Conclusion}
We have shown that the geometric structure describing the geometric phases in quantum control disturbed by classical stochastic processes is not the usual adiabatic principal bundle endowed with the Berry connection plus a deviation issued from the stochastic processes, but a composite bundle where the first floor represents the distortions of the ``control signals'' induced by the stochastic processes. The topology of the total manifold of this first floor plays an important role in the description. The total manifold of the second floor is an enlargement of the adiabatic bundle.

\end{document}